# Hydrogenated Aluminum Doped Zinc Oxide as Highly Transparent and Passivating Indium-Free Recombination Junction for TOPCon-Based Bottom Cell

Gökhan Altıner*[1], Jons Bolding[2], Yiğit Mert Kaplan[1], Floor Souren[2], Hindrik de Vries[2]
Raşit Turan[1,3], Hisham Nasser[1]

[1]ODTÜ-GÜNAM, Center for Solar Energy Research and Applications, 06800 Ankara, Türkiye
[2]SALD B.V., Hastelweg 257, 5652C, Eindhoven, Netherlands
[3]Department of Physics, Middle East Technical University, Ankara, 06800, Türkiye

*gokhan.altiner@odtugunam.org

## Abstract

Tandem solar cells offer a promising alternative to exceed the efficiency limits of single-junction silicon photovoltaics, yet they require high-performance recombination junctions that are transparent, passivating, and electrically efficient. Indium tin oxide (ITO), which is conventionally used as a recombination junction material, faces challenges related to indium scarcity and sputter-induced damage. This work investigates hydrogenated aluminum-doped zinc oxide (AZO:H) deposited by spatial atomic layer deposition (s-ALD) as a viable indium-free alternative for TOPCon-based bottom cells. The deposited AZO:H films demonstrate excellent transparency, exceeding 90% in the 380-1200 nm wavelength range. When applied to n-TOPCon surfaces with an $AlO_x$ capping layer, the stack achieves an outstanding passivation quality, indicated by implied open-circuit voltage ($iV_{oc}$) values up to 734 mV after annealing. The $AlO_x$ capping layer proved crucial for enhancing thermal stability by preventing hydrogen effusion at higher temperatures. While the contact resistivity was high for the 20 nm thick films tested, the combination of superior optical and passivation properties establishes spatial ALD-deposited AZO:H as a highly promising material for creating efficient and indium-free recombination junctions in next-generation tandem solar cells.

## 1. Introduction

The photovoltaic (PV) market is mostly dominated by single-junction crystalline silicon (c-Si) based solar cell technologies, due to c-Si being a reliable, mature, and non-toxic material. However, the efficiency of single-junction c-Si structures is theoretically limited to ~29% [1] and the recently reported single-junction c-Si solar cells have already reached an efficiency of 27.3% [2]. The theoretical limitation primarily stems from inefficient photon collection in single-junction cells: photons with energies lower than the bandgap of c-Si are not absorbed, while the higher energy photons will lead to thermalization-related losses. The most straightforward solution to increase the number of efficiently collected photons and surpass the Shockley-Queisser limit is combining different bandgap materials in a 2-terminal tandem structure, in which the top cell absorbs the short wavelength photons, and the bottom cell absorbs the high wavelength photons. These sub-cells are optically and electrically coupled with an interconnection junction. A favorable recombination junction needs to have high transparency to reduce the parasitic absorption, high passivation capabilities to passivate the surfaces of the sub cells, low contact resistivity, and high sheet resistance to palliate the effects of shunting [3]. ITO is a common choice for a recombination junction for its electrical and optical properties; however, indium is a scarce material and it is deposited by sputtering which damages the Si layer underneath [4]. AZO:H is a highly transparent and passivating indium-free alternative for a recombination junction. It shows high transparency and excellent passivation with fairly low contact resistivities [5]. While it is possible to deposit AZO:H by sputtering at room temperature [6], it is also possible to deposit AZO:H with various methods that do not damage the layers below, such as ALD.

## 2. Methods

AZO:H layers and AlO$_x$ capping layers were deposited using a Spatial ALD tool by SALD B.V. For the AZO:H deposition, diethylzinc (DEZ) and dimethylaluminum isopropoxide (DMAI) precursors were co-injected simultaneously with H$_2$O as co-reactant. DMAI was chosen over trimethylaluminum (TMA) as the doping precursor due to its superior doping efficiency [7], and it was also used for the deposition of AlO$_x$ capping layers. Different AZO:H deposition recipes were used varying the deposition temperature (110, 200, and 230°C) and H$_2$O co-reactant flow while keeping the DEZ and DMAI flow fixed. The AlO$_x$ capping layer recipe was kept fixed, but the deposition temperatures were changed so that both AZO:H and AlO$_x$ deposition temperatures would match for each sample. For the transmission-reflection measurements, 20 nm, 80 nm, and 200 nm AZO:H were deposited with and without an AlO$_x$ capping layer on 150 µm thick glass substrates. To investigate the passivation capability of AZO:H(/AlO$_x$), symmetrical 110 nm phosphorus-doped n-TOPCon were prepared on double-side textured Cz p-type c-Si wafers (~0.9 Ohm.cm, 170 µm). Later, 20 nm AZO:H(/AlO$_x$) layers were symmetrically deposited on the TOPCon test samples, and then the samples experienced 5 minutes of annealing under forming gas at different temperatures in the 375-525°C range. The thickness and optical properties of the deposited layers were extracted with variable angle spectroscopic ellipsometry (VASE). The *iV$_{oc}$* values were measured using photoconductance decay measurements (Sinton WCT-120 lifetime tester). Transmission-reflection results were measured by Bentham Instruments PVE300 characterization system with integrating sphere. The symmetric n-TOPCon/AZO:H samples were also used to extract the contact resistivity by screen printing low temperature metal contacts in the transfer length method (TLM) configuration.

## 3. Results

The transmission results for AZO:H with 20 nm (both with and without a capping layer) and 200 nm, along with the bare glass substrate, are given in Figure 1A. AZO:H on glass substrate demonstrates excellent transparency for 20 and 200 nm thick AZO:H resulting in more than 80% between 380-1200 nm wavelengths, while the 20 nm AZO:H predominantly maintains a transmission level of ~90%. The transmission of AZO:H for varying deposition temperatures (Figure 1B) shows that the AZO:H deposited at 110°C results in even higher transmission than that prepared at 200 and 230°C, reaching 90% at 550 nm wavelength.

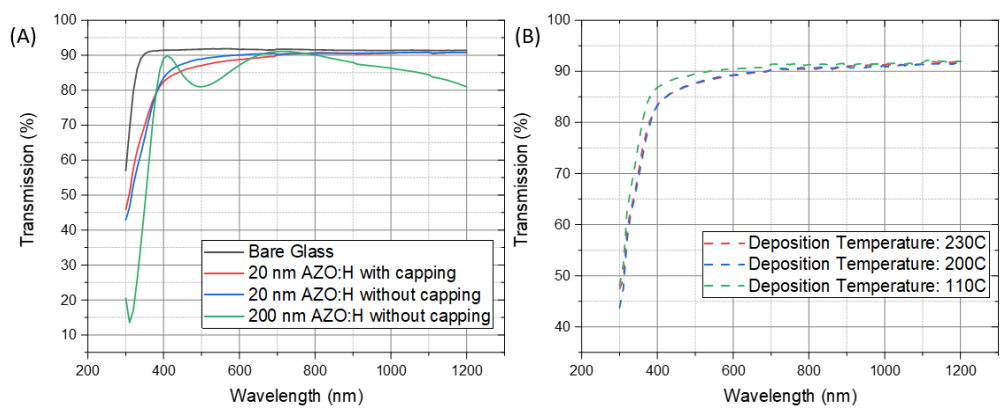

***Figure 1.*** *The transmission spectra of **(A)** different thicknesses of AZO:H with and without capping and the reference glass, **(B)** AZO:H deposited at varying temperatures.*

The change in $iV_{oc}$ values of the symmetrical n-TOPCon samples are given in Figure 2. TOPCon/AZO:H samples with $AlO_x$ showed outstanding passivation quality, peaking at 450 °C, with impressive $iV_{oc}$ values reaching 734 mV, demonstrating the outstanding capability of AZO:H/$AlO_x$ stacks as an excellent hydrogen source. While their uncapped counterparts were still able to improve the passivation by up to 56 mV, reaching 705 mV at 425°C, they resulted in lower $iV_{oc}$ values. In addition, uncapped samples showed lower thermal stability. At temperatures higher than 425°C, the passivation level of uncapped samples started decreasing, likely due to hydrogen effusion from uncapped AZO:H during annealing. This decrease is less evident for the samples with capping, especially at 525°C, since the capping layer partially prevents the effusion. The AZO:H deposited at 110°C outperformed the one deposited at 200°C and resulted in the largest improvement by about 90 mV. The 4-point-probe measurements on the glass substrates demonstrate that the sheet resistance of AZO:H strongly depends on their thickness with 20 nm AZO:H resulting in 300-650 kOhm/sq while the 200 nm ones give remarkably low values of ~90-170 Ohm/sq. The scanning electron microscope (SEM) images in Figure 3 show that the crystal size increases with increasing film thickness, which might explain the decreased sheet resistance of the thicker AZO:H samples. In Figure 3C, the cross-section view of the 200 nm AZO:H, deposited at 230°C with 40 mg/min $H_2O$ flow is given, and the vertical crystal growth direction is observed for the AZO:H which can be favorable for the application of ALD-AZO:H as a recombination junction.

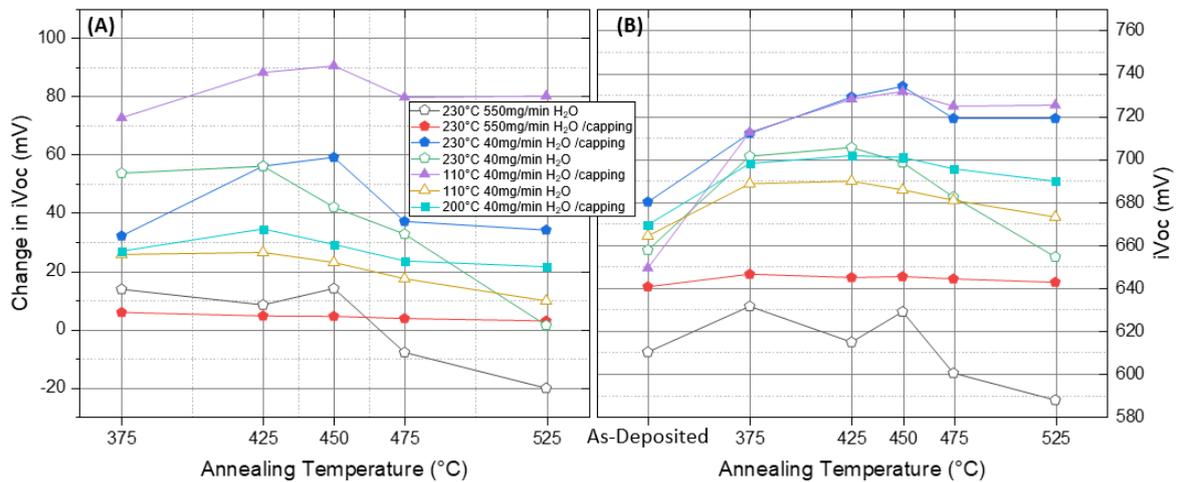

**Figure 2. (A)** Change in $iV_{oc}$ and **(B)** $iV_{oc}$ values for various AZO:H(/$AlO_x$) on symmetrical n-TOPCon samples.

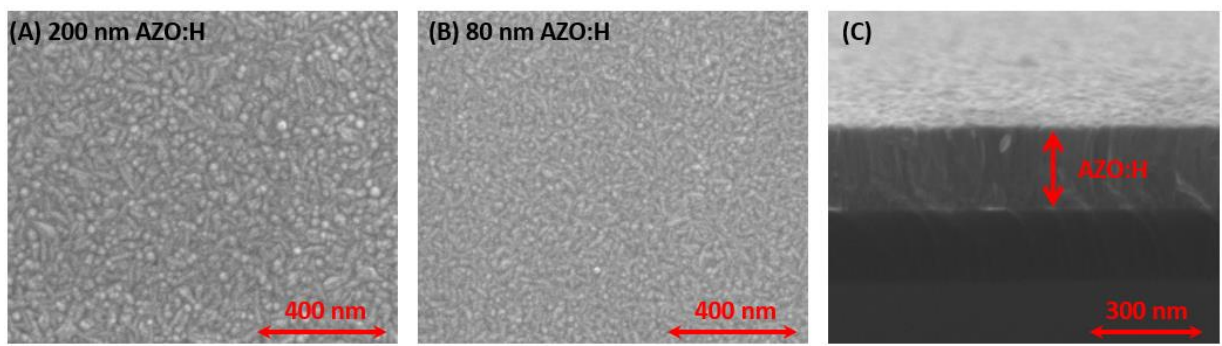

**Figure 3.** SEM image of **(A)** 200 nm AZO:H, **(B)** 80 nm AZO:H, and **(C)** the cross-section view of the 200 nm AZO:H.

The contact resistivity measurements give >180 mOhm.cm$^2$ for 20 nm AZO:H deposited at 230°C with 40 mg/min $H_2O$ flow on n-TOPCon and it shows unpronounced change after FGA. Although these results are considerably high, it is possible to decrease the contact resistivity by more than one order, without altering the passivation quality, by reducing the AZO:H thickness to 5 nm from 20 nm [5].

## 4. Conclusion

In this study, we successfully demonstrated the potential of AZO:H, deposited via spatial ALD, as a high-performance, indium-free recombination junction for TOPCon-based tandem solar cells. The results highlight the advantages of the material. Optically, the AZO:H films exhibit excellent transparency, which is critical for minimizing parasitic absorption and maximizing light transmission to the bottom cell. In terms of passivation, the AZO:H/AlO$_x$ stack delivered an outstanding *iV$_{oc}$* of 734 mV on n-TOPCon surfaces, underscoring its effectiveness as a hydrogen source for surface defect curing. The inclusion of the AlO$_x$ capping layer was shown to be essential for maintaining this high level of passivation across a wider thermal budget. Although the measured contact resistivity for 20 nm AZO:H films was high, established research indicates a clear pathway for optimization by reducing the film thickness without compromising passivation quality. Overall, the excellent transparency, superior passivation capabilities, and tunable electrical properties make AZO:H a compelling and promising candidate to replace conventional ITO in future high-efficiency silicon tandem solar cell architectures.

## 5. Acknowledgement

Funded by the European Union. Views and opinions expressed are however those of the author(s) only and do not necessarily reflect those of the European Union or RIA. Neither the European Union nor the granting authority can be held responsible for them. NEXUS project has received funding from the European Union's Horizon Europe research and innovation program under grant agreement No. 101075330.